# Classification and reconstruction of spatially overlapping phase images using diffractive optical networks

Deniz Mengu[a,b,c], Muhammed Veli[a,b,c], Yair Rivenson[a,b,c], Aydogan Ozcan*[abc]

[a]*Department of Electrical & Computer Engineering, University of California Los Angeles (UCLA), California, USA*

[b]*Department of Bioengineering, University of California Los Angeles (UCLA), California, USA*

[c]*California NanoSystems Institute (CNSI), University of California Los Angeles (UCLA), California, USA*

**E-mail: ozcan@ucla.edu*

**Abstract**

Diffractive optical networks unify wave optics and deep learning to all-optically compute a given machine learning or computational imaging task as the light propagates from the input to the output plane. Here, we report the design of diffractive optical networks for the classification and reconstruction of spatially overlapping, phase-encoded objects. When two different phase-only objects spatially overlap, the individual object functions are perturbed since their phase patterns are summed up. The retrieval of the underlying phase images from solely the overlapping phase distribution presents a challenging problem, the solution of which is generally not unique. We show that through a task-specific training process, passive diffractive networks composed of successive transmissive layers can all-optically and simultaneously classify two different randomly-selected, spatially overlapping phase images at the input. After trained with ~550 million unique combinations of phase-encoded handwritten digits from the MNIST dataset, our blind testing results reveal that the diffractive network achieves an accuracy of >85.8% for all-optical classification of two overlapping phase images of new handwritten digits. In addition to all-optical classification of overlapping phase objects, we also demonstrate the reconstruction of these phase images based on a shallow electronic neural network that uses the highly compressed output of the diffractive network as its input (with e.g., ~20-65 times less number of pixels) to rapidly reconstruct both of the phase images, despite their spatial overlap and related phase ambiguity. The presented phase image classification and reconstruction framework might find applications in e.g., computational imaging, microscopy and quantitative phase imaging fields.

## Introduction

Diffractive Deep Neural Networks ($D^2NN$) [1] have emerged as an optical machine learning framework that parameterizes a given inference or computational task as a function of the physical traits of a series of engineered surfaces/layers that are connected by diffraction of light. Based on a given task and the associated loss function, deep learning-based optimization is used to configure the transmission or reflection coefficients of the individual pixels/neurons of the diffractive layers so that the desired function is approximated in the optical domain through the light propagation between the input and output planes of the diffractive network [1–24]. Upon the convergence of this deep learning-based training phase using a computer, the resulting diffractive surfaces are fabricated using, e.g. 3D printing or lithography, to physically form the diffractive optical network which computes the desired task or inference, without the need for a power source, except for the illumination light.

A diffractive network can be considered as a coherent optical processor, where the input information can be encoded in the phase and/or amplitude channels of the sample/object field-of-view. Some of the previous demonstrations of diffractive networks utilized 3D printed diffracted layers operating at terahertz (THz) wavelengths to reveal that they can generalize to unseen data achieving >98% and >90% blind testing accuracies for amplitude-encoded handwritten digits (MNIST) and phase-encoded fashion products (Fashion-MNIST), respectively, using passive diffractive layers that collectively compute the all-optical inference at the output plane of the diffractive network [1,5,8]. In a recent work [6], diffractive optical networks have been utilized to all-optically infer the data classes of input objects that are illuminated by a broadband light source using only a single-pixel detector at the output plane. This work demonstrated that a broadband diffractive network can be trained to extract and encode the spatial features of input objects into the power spectrum of the diffracted light to all-optically reveal the object classes based on the spectrum of the incident light on a single-pixel detector. Deep learning-based training of diffractive optical networks have also been utilized in solving challenging inverse optical design problems e.g., ultra-short pulse shaping and spatially-controlled wavelength demultiplexing [3,4].

In general, coherent optical processing and the statistical inference capabilities of diffractive optical networks can be exploited to solve various inverse imaging and object classification problems through low-latency, low-power systems composed of passive diffractive layers. One such inverse problem arises when different phase objects reside on top of each other within the sample field-of-view of a coherent imaging platform: the spatial overlap between phase-only thin samples inevitably causes loss of spatial information due to the summation of the overlapping phase distributions describing the individual objects, hence, creating spatial phase ambiguity at the input field-of-view.

Here, we present phase image classification diffractive networks that can solve this phase ambiguity problem and simultaneously classify two spatially overlapping images through the same trained diffractive network (see Fig. 1). In order to address this challenging optical inference problem, we devised four alternative diffractive network designs (Figs. 1b-e) to all-optically infer the data classes of spatially overlapping phase objects. We numerically demonstrated the efficacy of these diffractive network designs in revealing the individual classes of overlapping phase objects using training and testing datasets that are generated based on phase-encoded MNIST digits [25]. Our diffractive optical networks were trained using ~550 *million* different input phase images containing spatially overlapping MNIST digits (from the same class as well as different classes); blind testing of one of the resulting diffractive networks using 10,000 test images of overlapping phase objects revealed a classification accuracy of >85.8%, optically matching the correct labels of both phase objects that were spatially overlapping within the input field-of-view.

In addition to all-optical classification of overlapping phase images using a diffractive network, we also combined our diffractive network models with separately trained, electronic image reconstruction networks to recover the individual phase images of the spatially overlapping input objects solely based on the optical class signals collected at the output of the corresponding diffractive network. We quantified the success of these digitally reconstructed phase images using the structural similarity index measure (SSIM) and the peak-signal-to-noise-ratio (PSNR) to reveal that a shallow electronic neural network with 2 hidden layers can simultaneously reconstruct both of the phase objects that are spatially overlapping at the input plane despite the fact that the number of detectors/pixels at the output plane of the diffractive network is e.g., ~20-65 times smaller compared to an ideal diffraction-limited imaging system. This means the diffractive network encoded the spatial features of the overlapping phase objects into a much smaller number of pixels at its output plane, which was successfully decoded by the shallow electronic network to simultaneously perform two tasks: (1) image reconstruction of overlapping spatial features at the input field-of-view, and (2) image decompression.

We believe that the presented diffractive network training and design techniques for computational imaging of phase objects will enable memory-efficient, low-power and high frame-rate alternatives to existing phase imaging platforms that often rely on high-pixel count sensor arrays, and therefore might find applications in e.g. microscopy and quantitative phase imaging fields.

**Results**

Spatial overlap between phase objects within the input field-of-view of an optical imaging system obscures the information of samples due to the superposition of the individual phase channels, leading to loss of structural information. For thin phase-only objects (such as e.g., cultured cells or thin tissue sections), when two samples $e^{j\theta_1(x,y)}$ and $e^{j\theta_2(x,y)}$ overlap with each other in space, the resulting object function can be expressed as $e^{j(\theta_1(x,y)+\theta_2(x,y))}$, and therefore a coherent optical imaging system does not have direct access to $\theta_1(x,y)$ or $\theta_2(x,y)$, except their summation (see Fig. 1a). In the context of diffractive networks and all-optical image classification tasks, another challenging aspect of dealing with spatially overlapping phase objects is that the effective number of data classes represented by different input images significantly increases compared to a single-object classification task. Specifically, for a target dataset with $M$ data classes represented through the phase channel of the input, the total number of data classes at the input (with two overlapping phase objects) becomes $C\binom{M}{2} + M = \frac{M(M-1)}{2} + M$, where $C$ refers to the combination operation . This means that if the diffractive optical network design assigns a single output detector to represent each one of these combinations, one would need $\frac{M(M-1)}{2} + M$ individual detectors. With the use of a differential detection scheme [5] that replaces each class detector with a pair of detectors (virtually representing the positive and negative signals), then the number of detectors at the output plane further increases to $2 \times \left(\frac{M(M-1)}{2} + M\right)$.

To mitigate this challenge, in this work we introduced different class encoding schemes that better handle the all-optical classification of these large number of possible class combinations at the input. The output detector layout, D-1, shown in Fig. 1b illustrates one alternative design strategy where the problem of classification of overlapping phase objects is solved by using only $2M$ individual detectors with a significant reduction in the number of output detectors when compared to $\frac{M(M-1)}{2} + M$. The use of $2M$ single-pixel detectors at the output plane (see Fig. 1b), can handle all the combinations and classify the overlapping input phase objects even if they belong to the same data class or not. To achieve this, we have two different sets of detectors, $\{D_m^i, m = 0,1,2, \dots, M-1\}$ and $\{D_m^{ii}, m = 0,1,2, \dots, M-1\}$, which represent the classes of the individual overlapping phase images. The final class assignments in this scheme are given based on the largest two optical signals among all the $2M$ detectors, where the assigned indices ($m$) of the corresponding two winner detectors indicate the all-optical classification results for the overlapping phase images. This is a simple class decision rule with a look up table of detector-class assignments (as shown Fig. 1b), where the strongest two detector signals indicate the inferred classes based on their $m$. Stated mathematically, the all-optical estimation of the classes, $\hat{\boldsymbol{c}} = [\widehat{c_1}, \widehat{c_2}]$, of the overlapping phase images is given by,

$$\hat{\boldsymbol{c}} = mod(argmax_2(\boldsymbol{I}), M) \qquad 1$$

where $\boldsymbol{I}$ denotes the optical signals detected by $2M$ individual detectors, i.e., $[D_m^i, D_m^{ii}]$. With the $mod(*)$ operation in Eq. 1, it can be observed that when the ground truth object classes, $c_1$ and $c_2$, are identical, a correct optical inference would result in $\widehat{c_1} = \widehat{c_2}$. On the other hand, when $c_1 \neq c_2$, there are four different detector combinations for the two largest optical signals that would result in the same $(\widehat{c_1}, \widehat{c_2})$ pair according to our class decision rule. For example, in the case of the input transmittance shown in Fig. 1a, which is comprised of handwritten digits '6' and '7', the output object classes based on our decision rule would be the same if the two largest optical signals collected by the detectors correspond to: (1) $D_6^i$ and $D_7^{ii}$, (2) $D_7^i$ and $D_6^{ii}$, (3) $D_6^i$ and $D_7^i$ or (4) $D_6^{ii}$ and $D_7^{ii}$; all of these four combinations of winner detectors at the output plane would reveal the correct classes for the input phase objects in this example (digits '6' and '7').

Therefore, the training the diffractive optical networks according to this class decision rule requires subtle but vital changes in the ground truth labels representing the inputs and the loss function driving the evolution of the diffractive layers compared to a single-object classification system. If we denote the one-hot vector labels representing the classes of the input objects in a single-object classification system as, $\boldsymbol{g^1}$ and $\boldsymbol{g^2}$, with an entry of 1 at their $c_1^{th}$ and $c_2^{th}$ entries, respectively, for the case of spatially overlapping two phase objects at the input field-of-view we can define new ground truth label vectors of length $2M$ using $\boldsymbol{g^1}$ and $\boldsymbol{g^2}$. For the simplest case of $c_1 = c_2$ (i.e., $\boldsymbol{g^1} = \boldsymbol{g^2}$), the $2M$-vector $\boldsymbol{g^e}$ is constructed as $\boldsymbol{g^e} = 0.5 \times [\boldsymbol{g^1}, \boldsymbol{g^2}]$. The constant multiplicative factor of 0.5 ensures that the resulting vector $\boldsymbol{g^e}$ defines a discrete probability density function satisfying $\sum_1^{2M} {g^e}_m = 1$. It is important to note that since $c_1 = c_2$, we have $[\boldsymbol{g^1}, \boldsymbol{g^2}] = [\boldsymbol{g^2}, \boldsymbol{g^1}]$. On the other hand, when the overlapping input phase objects are from different data classes i.e., $c_1 \neq c_2$, we define four different label vectors $\{\boldsymbol{g^a}, \boldsymbol{g^b}, \boldsymbol{g^c}, \boldsymbol{g^d}\}$ representing all the four combinations. Among this set of label vectors, we set $\boldsymbol{g^a} = 0.5 \times [\boldsymbol{g^1}, \boldsymbol{g^2}]$ and $\boldsymbol{g^b} = 0.5 \times [\boldsymbol{g^2}, \boldsymbol{g^1}]$. The label vectors $\boldsymbol{g^c}$ and $\boldsymbol{g^d}$ depict the cases, where the output detectors corresponding to the input object classes lie within $D_m^i$ and $D_m^{ii}$, respectively. In other words, the $c_1^{th}$ and $c_2^{th}$ entries of $\boldsymbol{g^c}$ are equal to 0.5, and similarly the $(M + c_1)^{th}$ and $(M + c_2)^{th}$ entries of $\boldsymbol{g^d}$ are equal to 0.5, while all the rest of the entries are equal to zero.

Based on these definitions, the training loss function ($\mathcal{L}$) of the associated forward model was selected to reflect all the possible input combinations at the sample field-of-view (input), therefore, it was defined as,

$$\mathcal{L} = (1 - |sgn(c_1 - c_2)|) \times \mathcal{L}_c^e + |sgn(c_1 - c_2)| \times \min\{\mathcal{L}_c^a, \mathcal{L}_c^b, \mathcal{L}_c^c, \mathcal{L}_c^d\} \qquad 2$$

where $\mathcal{L}_c^a, \mathcal{L}_c^b, \mathcal{L}_c^c$, $\mathcal{L}_c^d$, and $\mathcal{L}_c^e$ denote the penalty terms computed with respect to the ground truth label vectors $\boldsymbol{g^a}$, $\boldsymbol{g^b}$, $\boldsymbol{g^c}$, $\boldsymbol{g^d}$, and $\boldsymbol{g^e}$, respectively, and $sgn(.)$ is the signum function. The classification errors, $\mathcal{L}_c^x$, are computed using the cross-entropy loss [26],

$$\mathcal{L}_c^x = -\sum_{m=1}^{2M} g_m^x \log\left(\frac{e^{\overline{I_m}}}{\sum_{k=1}^{2M} e^{\overline{I_k}}}\right) \qquad 3$$

where *x* refers to one of *a*, *b*, *c*, *d*, or *e*, $\overline{I_m}$ denotes the normalized intensity collected by a given detector at the output plane (see the Methods section for further details). The term $g_m^x$ in Eq. (3) denotes the $m^{th}$ entry of the ground truth data class vector, $\boldsymbol{g^x}$.

Based on this diffractive network design scheme and the output detector layout D-1, we trained a 5-layer diffractive network (Figs. 1,a,b) using the loss function depicted in Eq. 2 over ~550 *million* input training images containing various combinations of spatially overlapping, phase-encoded MNIST handwritten digits. Following the training phase, the resulting diffractive layers of this network, which we term as $D^2$NN-D1, are illustrated in Fig. 2a. To quantify the generalization performance of $D^2$NN-D1 for the classification of overlapping phase objects that were never seen by the network before, we created a test dataset, $T_2$, with 10K phase images, where each image contains two spatially-overlapping phase-encoded test digits randomly selected from the standard MNIST test set, $T_1$. In this blind testing phase, $D^2$NN-D1 achieved 82.70% accuracy on $T_2$, meaning that in 8,270 cases out of 10,000 test inputs, the class estimates $[\widehat{c_1}, \widehat{c_2}]$ at the diffractive network's output plane were correct for both of the spatially overlapping handwritten digits. For the remaining 1730 test images, the classification decision of the diffractive network is incorrect for at least one of the phase objects within the field-of-view. Figures 2b-e depict some of the correctly classified phase image examples from the test dataset $T_2$ with phase encoded handwritten digits, along with the resulting class scores at the output detectors of the diffractive network.

This blind inference accuracy of the diffractive network shown in Fig. 2a, i.e., $D^2$NN-D1, can be further improved by combining the above outlined training strategy with a differential detection scheme, where each output detector in D1 (Fig. 1b) is replaced with a differential pair of detectors (i.e., a total of 2x2M detectors are located at the output plane, see Fig. 1c). The differential signal between a pair of detectors shown in Fig. 1c encodes a total of 2xM differential optical signals and similar to the previous approach of D1, the final class assignments in this scheme are given based on the two largest signals among all the differential optical signals. With the incorporation of this differential detection scheme, the vector $\boldsymbol{I}$ in Eq. 1 is replaced with the differential signal [5], $\Delta\boldsymbol{I} = \boldsymbol{I}_+ - \boldsymbol{I}_-$, where $\boldsymbol{I}_+$ and $\boldsymbol{I}_-$ denote the optical signals collected by the $2M$ detector pairs, virtually representing the positive and negative parts, respectively.

Using this differential diffractive network design, which we termed as $D^2$NN-D1d (see Fig. 1c), we achieved a blind testing accuracy of 85.82% on the test dataset $T_2$. The diffractive layers comprising the $D^2$NN-D1d network are shown in Fig. 3a, which were trained using ~$550\ million$ input phase images of spatially overlapping MNIST handwritten digits, similar to $D^2$NN-D1. Compared to the classification accuracy attained by $D^2$NN-D1, the inference accuracy of its differential counterpart, $D^2$NN-D1d, is improved by >3.1% at the expense of using $2M$ additional detectors at the output plane of the optical network. Figures 3b-d illustrate some examples of the correctly classified phase images from the test dataset $T_2$ with phase encoded handwritten digits, along with the resulting differential class scores at the output detectors of the diffractive network.

The blind inference accuracies achieved by $D^2$NN-D1 and $D^2$NN-D1d (82.70% and 85.82%, respectively) on the test dataset $T_2$, demonstrate the success of the underlying detector layout designs and the associated training strategy for solving the phase ambiguity problem to all-optically classify overlapping phase images using diffractive networks. When these two diffractive optical networks ($D^2$NN-D1 and $D^2$NN-D1d) are blindly tested over $T_1$ that provides input images containing a single phase-encoded handwritten digit (without the second overlapping phase object), they attain better classification accuracies of 90.59% and 93.30%, respectively (see the Methods section). As a reference point, a 5-layer diffractive network design with an identical layout to the one shown in Fig. 1a, can achieve a blind classification accuracy of ~98% [5,8] on test set $T_1$, provided that it is trained to classify only one phase-encoded handwritten digit per input image (without any spatial overlap with other objects). This reduced classification accuracy of $D^2$NN-D1 and $D^2$NN-D1d on test set $T_1$ (when compared to ~98%) indicates that their forward training model, driven by the loss functions depicted in Eqs. 2-3, guided the evolution of the corresponding diffractive layers to recognize the spatial features created by the overlapping handwritten digits, as opposed to focusing solely on the actual features describing the individual handwritten digits.

To further reduce the required number of optical detectors at the output plane of a diffractive network, we considered an alternative design (D-2) shown in Fig. 1d. In this alternative design scheme D-2, there are two extra detectors $\{D_Q^+, D_Q^-\}$ (shown with blue in Fig. 1d), in addition to $M$ class detectors $\{D_m, m = 1,2, \dots, M\}$ (shown with gray in Fig. 1d). The sole function of the additional pair of detectors $\{D_Q^+, D_Q^-\}$ is to decide whether the spatially-overlapping input phase objects belong to the same or different data classes. If the difference signal of this differential detector pair (Fig. 1d) is non-negative (i.e., $I_{D_Q^+} \geq I_{D_Q^-}$), the diffractive network will infer that the overlapping input objects are from the same data class, hence there is only one class assignment to be made by simply determining the maximum signal at the output class detectors: $\{D_m, m = 1,2, \dots, M\}$. A negative signal difference between $\{D_Q^+, D_Q^-\}$, on the other hand, indicates that the two overlapping phase objects are from different data classes/digits, and the final class assignments in this case of $I_{D_Q^+} < I_{D_Q^-}$ are given based on the largest two optical signals among all the remaining $M$ detectors at the network output, $\{D_m, m = 1,2, \dots, M\}$. Refer to the Methods section for further details on the training of diffractive networks that employ D-2 (Fig. 1d)

Similar to earlier diffractive network designs, we used ~550 $million$ input phase images of spatially overlapping MNIST handwritten digits to train 5 diffractive layers constituting the $D^2$NN-D2 network (see Fig. 4a). Figures 4b-d illustrate sample input phase images that contain objects from different data classes, along with the output detector signals that correctly predict the classes/digits of these overlapping phase objects; notice that in each one of these cases, we have at the output plane $I_{D_Q^+} < I_{D_Q^-}$ indicating the success of the network's inference. As another example of blind testing, Fig. 4e reports the diffractive network's inference for two input phase objects that are from the same data class, i.e., digit '3'. At the network's output, this time we have $I_{D_Q^+} > I_{D_Q^-}$, correctly predicting that the two overlapping phase images are of the same class; the maximum output signal of the remaining output detectors $\{D_m, m = 1,2, \dots, M\}$ also correctly reveals that the handwritten phase images belong to digit '3' with a maximum signal at $D_3$. This $D^2$NN-D2 design provides 82.61% inference accuracy on the test set $T_2$ with 10K test images, closely matching the inference performance of $D^2$NN-D1 (82.70%) reported in Fig 2. In fact, an advantage of this $D^2$NN-D2 design lies in its inference performance and blind testing accuracy on test set $T_1$, achieving 93.38% for classification of input phase images of single digits (without any spatial overlap at the input field-of-view).

We also implemented the differential counterpart of the detector layout D-2, which we term as D-2d (see Fig. 1e), where the $M$ class detectors in D-2 are replaced with $M$ differential pairs of output detectors. In this configuration D-2d, the total number of detectors at the output plane of the diffractive network becomes $2M + 2$ and the all-optical inference rules remain the same as in D-2: for $I_{D_Q^+} \geq I_{D_Q^-}$, the class inference is made by simply determining the maximum differential signal at the output class detectors, and for the case of $I_{D_Q^+} < I_{D_Q^-}$ the inference of the classes of input phase images is determined based on the largest two differential optical signals at the network output. Figure 5a shows the diffractive layers of the resulting $D^2$NN-D2d that is trained based on the detector layout, D-2d (Fig. 1e) using the same training dataset as before: ~550 $million$ input phase images of spatially overlapping, phase-encoded MNIST handwritten digits. This new differential diffractive network design, D$^2$NN-D2d, provides significantly higher blind inference accuracies compared to its non-differential counterpart D$^2$NN-D2, achieving 85.22% and 94.20% on $T_2$ and $T_1$ datasets, respectively. Figures 5b-d demonstrate some examples of the input phase images from test set $T_2$ that are correctly classified by D$^2$NN-D2d along with the corresponding optical signals collected by the output detectors representing the positive and negative parts, $\boldsymbol{I_{M+}}$ and $\boldsymbol{I_{M-}}$, of the associated differential class signals, $\Delta\boldsymbol{I}_M = \boldsymbol{I_{M+}} - \boldsymbol{I_{M-}}$. As another example, the input phase image depicted in Fig. 5e has two overlapping phase-encoded digits from the same data class, handwritten digit ‘4’, and the diffractive optical network correctly outputs $I_{D_Q^+} > I_{D_Q^-}$ with the maximum differential class score strongly revealing an optical inference of digit ‘4’.

Table 1 summarizes the optical blind classification accuracies achieved by different diffractive optical network designs, D$^2$NN-D1, D$^2$NN-D1d, D$^2$NN-D2 and D$^2$NN-D2d on test image sets $T_2$ and $T_1$. Even though D$^2$NN-D1d achieves the highest inference accuracy for the classification of spatially overlapping phase objects, D$^2$NN-D2d offers a balanced optical inference system achieving very good accuracy on both $T_1$ and $T_2$. These two differential diffractive network models outperform their non-differential counterparts with superior inference performance on both $T_2$ and $T_1$. The confusion matrices demonstrating the class-specific inference performances of the presented diffractive networks, D$^2$NN-D1, D$^2$NN-D1d, D$^2$NN-D2 and D$^2$NN-D2d, are also reported in Supplementary Figures 1-4, respectively.

Next, we aimed to reconstruct the individual images of the overlapping phase objects (handwritten digits) using the detector signals at the output of a diffractive network; stated differently our goal here is to resolve the phase ambiguity at the input plane and reconstruct both of the input phase images, despite their spatial overlap and the loss of phase information. For this aim, we combined each one of our diffractive optical networks, D$^2$NN-D1, D$^2$NN-D1d, D$^2$NN-D2 and D$^2$NN-D2d, one by one, with a shallow, fully-connected (FC) electronic network with two hidden layers, forming a task-specific imaging system as shown in Fig. 6. In these hybrid machine vision systems, the optical signals synthesized by a given diffractive network (front-end encoder) are interpreted as encoded representations of the spatial information content at the input plane. Accordingly, the electronic back-end neural network is trained to process the encoded optical signals collected by the output detectors of the diffractive network to decode and reconstruct the individual phase images describing each object function at the input plane, resolving the phase ambiguity due to the spatial overlap of the two phase objects. Figures 6a-d illustrate 3 different input images taken from the test set $T_2$ for each diffractive network design ($D^2$NN-D1, $D^2$NN-D1d, $D^2$NN-D2 and $D^2$NN-D2d) along with the corresponding image reconstructions at the output of each one of the electronic networks that are separately trained to work with the optical diffractive networks (front-end). As depicted in Fig. 6, the electronic image reconstruction networks only have 2 hidden layers with 100 and 400 neurons, and the final output layers of these networks have 28×28×2 neurons, revealing the images of the individual phase objects, resolving the phase ambiguity due to the spatial overlap of the input phase images. The quality of these image reconstructions is quantified using (1) the structural similarity index measure (SSIM) and (2) the peak signal-to-noise ratio (PSNR). Table 2 shows the mean SSIM and PSNR values achieved by these hybrid machine vision systems along with the corresponding standard deviations computed over the entire 10K test images ($T_2$). For these presented image reconstructions, we should emphasize that the dimensionality reduction (i.e., the image data compression) between the input and output planes of the diffractive networks ($D^2$NN-D1, $D^2$NN-D1d, $D^2$NN-D2 and $D^2$NN-D2d) is 39.2×, 19.6×, 65.33× and 35.63×, respectively, meaning that the spatial information of the overlapping phase images at the input field-of-view is significantly compressed (in terms of the number of pixels) at the output plane of the diffractive network. This large compression sets another significant challenge for the image reconstruction task in addition to the phase ambiguity and spatial overlap of the target images. With these large compression ratios, the presented diffractive network-based machine vision systems managed to faithfully recover the phase images of each input object despite their spatial overlap and phase information loss, demonstrating the coherent processing power of diffractive optical networks as well as the unique design opportunities enabled by their collaboration with electronic neural networks that form task-specific back-end processors.

## Discussion

The optical classification of overlapping phase images using diffractive networks presents a challenging problem due to the spatial overlap of the input images and the associated loss of phase information at the input plane. Interestingly, different combinations of handwritten digits at the input present different amounts of spatial overlap, which is a function of the class of the selected input digits as well as the style of the handwriting of the person. To shed more light on this, we quantified the all-optical blind inference accuracies of the presented diffractive optical networks as a function of the spatial overlap percentage, $\xi$, at the input field-of-view; see Fig. 7. In the first group of examples shown in Fig. 7a, the input fields-of-view contain digits from different data classes ($c_1 \neq c_2$) and in the second group of examples shown in Fig. 7b, the spatially overlapping objects are from the same data class, $c_1 = c_2$. The input phase images in $T_2$ exhibit spatial overlap percentages varying between ~20% and ~100%. Figures 7c,d illustrate the change in the optical blind inference accuracy of the diffractive network, $D^2$NN-D1, as a function of the spatial overlap percentage, $\xi$, for the first ($c_1 \neq c_2$) and the second ($c_1 = c_2$) group of test input images, respectively. When $c_1 \neq c_2$ as in Fig. 7c, the optical inference accuracy is hindered by the increasing amount of spatial overlap between the two input phase objects, as in this case, the spatial features of the effective input transmittance function significantly deviate from the features defining the individual data classes. In the other case shown in Fig. 7d, i.e., $c_1 = c_2$, the relationship between the spatial overlap ratio $\xi$ and the blind inference accuracy is reversed, since, with $c_1 = c_2$, increasing $\xi$ means that the effective phase distribution at the input plane resembles more closely to a single object/digit. The same behavior can also be observed for the other diffractive optical networks, $D^2$NN-D1d, $D^2$NN-D2 and $D^2$NN-D2d, reported in Figs. 7e-f, 7g-h, 7i-j, respectively.

In summary, to the best of our knowledge, this manuscript reports the first all-optical multi-object classification designs based on diffractive networks demonstrating their potential in solving challenging classification and computational imaging tasks in a resource-efficient manner using only a handful detectors at the output plane. In the context of optical classification and reconstruction of overlapping phase objects, also resolving the phase ambiguity due to the spatial overlap of input images, this study exclusively focuses on a setting where the thin input objects are only modulating the phase of the incoming waves, and absorption is negligible. Without loss of generality, the presented diffractive design schemes with the associated loss functions and training methods can also be extended to applications, where the input objects partially absorb the incoming light.

## Methods

## Optical Forward Model of Diffractive Networks

D$^2$NN framework formulates a given machine learning e.g., object classification or inverse design task as an optical function approximation problem and parameterizes that function over the physical features of the materials inside a diffractive black-box. As is the case in this study, this optical black-box is often modeled through a series of thin modulation layers connected by the diffraction of light waves. Here, we focused our efforts on 5-layer diffractive optical networks as shown in Fig. 1a, each occupying an area of $106\lambda \times 106\lambda$ on the lateral space with λ denoting the wavelength of the illumination light. The modulation function of each diffractive layer was sampled and represented over a 2D regular grid with a period of 0.53λ resulting in $N = 200 \times 200$ different transmittance coefficients, i.e., the diffractive 'neurons'. Based on the 0.53λ diffractive feature size, we set the layer-to-layer axial distance to be 40λ to ensure connectivity between all the neurons on two successive layers.

We selected the diffractive layer thickness, $h$, as a trainable physical parameter dictating the transmittance of each neuron together with the refractive index of the material. To limit the material thickness range during the deep learning-based training, $h$ is defined as a function of an auxiliary, learnable variable, $h_a$, and a constant base thickness, $h_b$,

$$h = Q_4(\frac{\sin(h_a) + 1}{2}(h_m - h_b)) + h_b \qquad 4,$$

where the function, $Q_*(.)$ represents the *-bit quantization operator and $h_m$ is the maximum allowed material thickness. If the material thickness over the $i^{th}$ diffractive neuron located at $(x_i, y_i, z_i)$ is denoted by $h(x_i, y_i, z_i)$, then the resulting transmittance coefficient of that neuron, $t(x_i, y_i, z_i)$, is given by,

$$t(x_i, y_i, z_i) = \exp\left(-\frac{2\pi\kappa(\lambda)h(x_i, y_i, z_i)}{\lambda}\right)\exp\left(j(n(\lambda) - n_s)\frac{2\pi h(x_i, y_i, z_i)}{\lambda}\right) \qquad 5,$$

where $n(\lambda)$ and $\kappa(\lambda)$ are the real and imaginary parts of complex-valued refractive index of the diffractive material at wavelength, $\lambda$. Following the earlier experimental demonstrations of diffractive optical networks, in this work we set the $n(\lambda)$ and $\kappa(\lambda)$ values to be 1.7227 and 0.031, respectively [8]. The parameter $n_s$ in Eq. 5 refers to the refractive index of the medium, surrounding the diffractive layers; without loss of generality, we assumed $n_s = 1$ (air). Based on the outlined material properties, the $h_m$ and $h_b$ in Eq. 4 were selected as 2λ and 0.66λ, respectively, ensuring that the phase modulation term in Eq. 5, $(n(\lambda) - n_s)\frac{2\pi h(x_i,y_i,z_i)}{\lambda}$, can cover the entire [0-2π] phase modulation range per diffractive feature/neuron.

Based on the Rayleigh-Sommerfeld diffraction integral, a neuron located at $(x_i, y_i, z_i)$ can be viewed as the source of a secondary wave,

$$w_i(x,y,z) = \frac{z - z_i}{r^2}\left(\frac{1}{2\pi r} + \frac{n_s}{j\lambda}\right)\exp\left(\frac{j2\pi n_s r}{\lambda}\right) \qquad 6,$$

where $r$ denotes the radial distance $\sqrt{(x-x_i)^2 + (y-y_i)^2 + (z-z_i)^2}$. Based on this, the output wave emanating from the $i^{th}$ neuron on the $k^{th}$ layer, $u_i^k(x,y,z)$ can be written as,

$$u_i^k(x,y,z) = w_i(x,y,z)t(x_i,y_i,z_i)\sum_{q=1}^{N} u_q^{k-1}(x_i,y_i,z_i) \qquad 7.$$

The term $\sum_{q=1}^{N} u_q^{k-1}(x_i,y_i,z_i)$ in Eq. 9 represents the wave incident on the $i^{th}$ neuron on the $k^{th}$ layer, generated by the neurons on the previous, $(k-1)^{th}$ diffractive layer.

In this study we also assumed that the transmittance function inside the input field-of-view, $T_{in}(x,y)$, covers an area of $53\lambda \times 53\lambda$ and without loss of generality, it is illuminated with a uniform plane wave. At the output plane, the width of each single-pixel detector was set to be $6.36\lambda$ on both x and y directions for all four output detector configurations shown in Figs. 1b-d. Based on the outlined optical forward model, the diffractive optical networks process the incoming waves generated by the complex-valued transmittance function, $T_{in}(x,y)$, formed by the overlapping thin phase objects and synthesize a 2D intensity distribution at the output plane for all-optical inference of the classes of the overlapping objects. The optical intensity distribution within the active area of each output detector is integrated to form elements of the vector, $\boldsymbol{I}$ in Eq. 1. The number of elements in this optical signal vector, $\boldsymbol{I}$, is equal to the number of output detectors, thus its length is $2M$, $4M$, $M+2$ and $2M+2$ for $D^2$NN-D1, $D^2$NN-D1d, $D^2$NN-D2 and $D^2$NN-D2d, respectively. As part of our forward training model, $\boldsymbol{I}$ is normalized to form, $\bar{\boldsymbol{I}}$,

$$\bar{\boldsymbol{I}} = \frac{\boldsymbol{I}}{c\max\{\boldsymbol{I}\}} \qquad 8.$$

where the coefficient $c$ in Eq. 8 serves as the temperature parameter of the softmax function depicted in Eq. 3, and it was empirically set to be 0.1 for training of all the diffractive networks. It is important to note that this normalization step in Eq. 8 is only used during the training stage, and once the training is finished, the forward inference directly uses the detected intensities to decide on the object classes based on the corresponding decision rules. While the vector $\bar{\boldsymbol{I}}$ is directly used in Eq. 3 for the D$^2$NN-D1 network, in the case of D$^2$NN-D1d, $\bar{\boldsymbol{I}}$ is split into two vectors of length $2M$, i.e., $\boldsymbol{I}_+$ and $\boldsymbol{I}_-$, representing the signals collected by the positive and negative detectors, and the associated differential signal is computed as $\Delta\boldsymbol{I} = \boldsymbol{I}_+ - \boldsymbol{I}_-$. Accordingly, during the training of D$^2$NN-D1d, the loss function depicted in Eq. 3, were computed using $\Delta\boldsymbol{I}$ instead of $\bar{\boldsymbol{I}}$.

For the diffractive network D$^2$NN-D2, the output of the normalization defined in Eq. 8 was split into two: $\boldsymbol{I_M}$ and $\boldsymbol{I_Q}$. The first part, $\boldsymbol{I_M}$, represents the optical signals coming from the $M$ class specific detectors in the detector layout D-2 (the gray detectors Fig. 1d). The latter, $\boldsymbol{I_Q}$, contains two entries describing the positive and the negative parts of the indicator signals, $I_{D_Q^+}$ and $I_{D_Q^-}$ (see the blue detectors Fig. 1d). These two extra detectors, $\{D_Q^+, D_Q^-\}$, form a differential pair that controls the functional form of the class decision rule based on the sign of the difference between the optical signals collected by these detectors. We accordingly determine the class assignments as follows,

$$\hat{\boldsymbol{c}} = \begin{Bmatrix} [argmax(\boldsymbol{I_M}), argmax(\boldsymbol{I_M})], & if\ I_{D_Q^+} \geq I_{D_Q^-} \\ argmax_2(\boldsymbol{I_M}), & otherwise. \end{Bmatrix} \quad 9$$

To enable the training of diffractive optical networks according to the class assignment rule in Eq. 9, we defined a loss function, $\mathcal{L} = \mathcal{L}_Q + \mathcal{L}_c$, that corresponds to the superposition of two different penalty terms, $\mathcal{L}_Q$ and $\mathcal{L}_c$. Here, $\mathcal{L}_Q$ represents the error computed with respect to the binary ground truth indicator signal, $g_Q$,

$$g_Q = \begin{Bmatrix} 1, & if\ c_1 = c_2 \\ 0, & otherwise \end{Bmatrix}. \quad 10$$

Accordingly, $\mathcal{L}_Q$ was defined as,

$$\mathcal{L}_Q = -g_Q \log\left(\sigma\left(I_{D_Q^+} - I_{D_Q^-}\right)\right) - \left(1 - g_Q\right)\log\left(1 - \sigma\left(I_{D_Q^+} - I_{D_Q^-}\right)\right) \quad 11$$

where $\sigma(\cdot)$ denotes the sigmoid function. The classification loss, $\mathcal{L}_c$, on the other hand, is identical to the cross-entropy loss depicted in Eq. 3, except that the vector $\boldsymbol{I}$ is replaced with $\boldsymbol{I_M}$. Unlike the previous diffractive networks (D$^2$NN-D1 and D$^2$NN-D1d), the forward model of the diffractive optical networks trained based on the output detector layout D-2 do *not* require multiple ground truth vector labels. Simply the ground truth label vector of a given input field-of-view is defined as $\boldsymbol{g} = \frac{\boldsymbol{g^1}+\boldsymbol{g^2}}{\mathbf{2}}$ satisfying the condition, $\sum_1^M g_m = 1$.

In the case of D[2]NN-D2d, the vector $\bar{\boldsymbol{I}}$ contains 3 main parts, $\boldsymbol{I_{M+}}$, $\boldsymbol{I_{M-}}$ and $\boldsymbol{I_Q}$ where $\boldsymbol{I_{M+}}$ and $\boldsymbol{I_{M-}}$ are length $M$ vectors containing the normalized optical signals collected by the detectors representing the positive and negative parts of the final differential class scores $\Delta\boldsymbol{I}_M = \boldsymbol{I_{M+}} - \boldsymbol{I_{M-}}$. Accordingly, in the associated forward training model, the intensity vector $\boldsymbol{I_M}$ in Eq. (9) is replaced with the differential signal, $\Delta\boldsymbol{I}_M$.

**Testing of diffractive optical networks on the dataset $T_1$**

During the blind testing of the presented diffractive optical networks on the test set $T_1$, the class estimation solely uses the $argmax$ operation, searching for the highest class-score synthesized by the diffractive networks, based on the associated output plane detector layouts shown in Fig. 1. The purpose of this performance quantification using $T_1$ is to reveal whether the diffractive optical networks trained based on overlapping input phase objects can learn and automatically recognize the characteristic spatial features of the individual handwritten digits (without any spatial overlap). For this goal, in the case of D[2]NN-D1 and D[2]NN-D1d, the class estimation rule in Eq. 1 was replaced with, $mod(argmax(\boldsymbol{I}), M)$ and $mod(argmax(\boldsymbol{\Delta I}), M)$, respectively. Since the input images in the test set $T_1$ contain single, phase-encoded handwritten digits without the second overlapping phase object, the optical signals collected by the detectors, $\{D_Q^+, D_Q^-\}$, at the output plane of the diffractive networks D[2]NN-D2 and D[2]NN-D2d become irrelevant for the classification of the images in $T_1$. Therefore, the decision rule in Eq. 9 is simplified to $argmax(\boldsymbol{I_M})$ and $argmax(\Delta\boldsymbol{I}_M)$ for the all-optical classification of the input test images in $T_1$ based on the diffractive networks D[2]NN-D2 and D[2]NN-D2d, respectively.

**Architecture and Training of the Phase Image Reconstruction Network**

The phase image reconstruction electronic neural networks (back-end) following each of the presented diffractive optical networks (front-end) include 4 neural layers. The number of neurons on their first layer is equal to the number of detectors at the output plane of the preceding diffractive network (D-1, D-1d, D-2 or D-2d, see Figs. 1a-d). The number of neurons, on the subsequent 3 layers are 100, 400, 1568, respectively. Note that the output layer of each image reconstruction electronic neural network has 2×28×28 neurons as it simultaneously reconstructs both of the overlapping phase objects, resolving the phase ambiguity due to the spatial overlap at the input plane. Each fully-connected (FC) layer constituting these image reconstruction networks applies the following operations:

$$\boldsymbol{\rho}_{\boldsymbol{l+1}} = \mathrm{BN}\{\mathrm{LReLU}[\mathrm{FC}\{\boldsymbol{\rho}_{\boldsymbol{l}}\}]\} \tag{12.}$$

where $\boldsymbol{\rho}_{\boldsymbol{l+1}}$ and $\boldsymbol{\rho}_{\boldsymbol{l}}$ denote the output and input values of the $l^{th}$ layer of the electronic network, respectively. The operator LReLU stands for the leaky rectified linear unit:

$$LReLU[\boldsymbol{x}] = \begin{Bmatrix} \boldsymbol{x}, & if\ x \geq 0 \\ 0.1\boldsymbol{x}, & otherwise. \end{Bmatrix}. \tag{13}$$

The batch normalization, BN, normalizes the activations at the output of LReLU to zero mean and a standard deviation of 1, and then shifts the mean to a new center, $\beta^{(l)}$, and re-scales it with a multiplicative factor, $\gamma^{(l)}$, where $\beta^{(l)}$ and $\gamma^{(l)}$ are learnable parameters, i.e.,

$$\mathrm{BN}[\boldsymbol{x}] = \gamma^{(l)} \cdot \frac{x - \mu_B^{(l)}}{\sqrt{{\sigma_B^{(l)}}^2 + \epsilon}} + \beta^{(l)} \tag{14a}$$

$$\mu_B = \frac{1}{m}\sum_{i=1}^{m} x_i, \quad \sigma_B^2 = \frac{1}{m}\sum_{i=1}^{m}(x_i - \mu_B)^2 \tag{14b}$$

The hyperparameter $\epsilon$ is a small constant that avoids division by 0 and it was taken as $10^{-3}$.

The training of the phase image reconstruction networks was driven by the reversed Huber (or "BerHu") loss, which computes the error between two images, $a(x,y)$ and $b(x,y)$, as follows:

$$BerHu(\boldsymbol{a},\boldsymbol{b}) = \sum_{\substack{x,y \\ |a(x,y)-b(x,y)| \leq \varphi}} |a(x,y) - b(x,y)| + \sum_{\substack{x,y \\ |a(x,y)-b(x,y)| > \varphi}} \frac{[a(x,y) - b(x,y)]^2 + \varphi^2}{2\varphi} \tag{15}$$

The hyperparameter $\varphi$ in Eq. 15 is a threshold for the transition between mean-absolute-error and mean-squared-error, and it was set to be 20% of the standard deviation of the ground truth image.

If we let $\phi_p(x,y)$ and $\phi_q(x,y)$ denote the first and second output of each image reconstruction electronic network (i.e., 28x28 pixels per phase object), the image reconstruction loss, $\mathcal{L}_r$, was defined as the *minimum* of two error terms, $\mathcal{L}_r'$ and $\mathcal{L}_r''$, i.e.,

$$\mathcal{L}_r = \min\{\mathcal{L}_r', \mathcal{L}_r''\}, \quad 16a$$

$$\mathcal{L}_r' = \frac{\left[\mathrm{BerHu}\left(\phi_p(x,y), \phi_1(x,y)\right) + \mathrm{BerHu}\left(\phi_q(x,y), \phi_2(x,y)\right)\right]}{2}, \quad 16b$$

$$\mathcal{L}_r'' = \frac{\left[\mathrm{BerHu}\left(\phi_p(x,y), \phi_2(x,y)\right) + \mathrm{BerHu}\left(\phi_q(x,y), \phi_1(x,y)\right)\right]}{2}, \quad 16c$$

where $\phi_1(x,y)$ and $\phi_2(x,y)$ denote the ground truth phase images of the first and second objects, respectively, which overlap at the input plane of the diffractive network. As there is no hierarchy or priority difference between the input objects $\phi_1(x,y)$ and $\phi_2(x,y)$, Eq. 16 lets the image reconstruction network to choose their order regarding its output activations.

**Other Details of Diffractive Network Training**

With the 0.53$\lambda$ lateral sampling rate in our forward optical model, the transmittance function inside the field-of-view, $T_{in}(x,y)$, was represented as a $100 \times 100$ discrete signal. In our diffractive network training, the 8-bit grayscale values of the MNIST digits were first converted to 32-bit double format, normalized to the range [0,1] and then resized to $100 \times 100$ using bilinear interpolation. If we denote these normalized and resized grayscale values of the two input objects/digits that overlap at the input plane as $\theta_1(x,y)$ and $\theta_2(x,y)$ then the transmittance function within the input field-of-view, $T_{in}(x,y)$, is defined as,

$$T_{in}(x,y) = e^{j\pi\theta_1(x,y)} e^{j\pi\theta_2(x,y)} \quad 17.$$

During the training of the presented diffractive networks, $\theta_1(x,y)$ and $\theta_2(x,y)$ are randomly selected from the standard 55K training samples of MNIST dataset without replacement, meaning that, an already selected training digit was not selected again until all 55K samples are depleted constituting an epoch of the training phase. In this manner, we trained the diffractive optical networks for 20,000 epochs, showing each optical network approximately *550 million* different $T_{in}(x,y)$ during the training phase. Supplementary Fig. 5 illustrates the convergence curves of our best performing diffractive networks models ($D^2$NN-D1d and $D^2$NN-D2d) over the course of these 20,000 epochs. Similarly, to generate the input fields in test dataset $T_2$, we randomly selected $\theta_1(x,y)$ and $\theta_2(x,y)$ among the standard 10K test samples of MNIST dataset, without replacement, and this was repeated two times providing us the 10K unique phase input images of overlapping handwritten digits constituting $T_2$. In our $T_2$ test set, 8998 inputs contain overlapping digits from different data classes, while the remaining 1002 inputs have overlapping samples from the same data class/digit. The validation image set (Supplementary Fig. 5), on the other hand, contains 5K unique phase input images created by randomly selecting $\theta_1(x,y)$ and $\theta_2(x,y)$ among the standard 10K test samples of MNIST dataset without replacement. Note that the $\theta_1(x,y)$ and $\theta_2(x,y)$ combinations used in the validation set of Supplementary Fig. 5 are not included in $T_2$ to achieve true blind testing without any data contamination.

The overlap percentage, $\xi$, between any given pair of samples, $\theta_1(x,y)$ and $\theta_2(x,y)$ (see Fig. 7), is quantified by,

$$\xi_1 = \frac{\sum_q \sum_p \left| sgn\left(\theta_2(x_q, y_p)\right)\right| \theta_1(x_q, y_p)}{\sum_{q'} \sum_{p'} \theta_1(x_{q'}, y_{p'})} \times 100 \qquad 18a,$$

$$\xi_2 = \frac{\sum_q \sum_p \left| sgn\left(\theta_1(x_q, y_p)\right)\right| \theta_2(x_q, y_p)}{\sum_{q'} \sum_{p'} \theta_2(x_{q'}, y_{p'})} \times 100 \qquad 18b,$$

$$\xi = \max\{\xi_1, \xi_2\} \qquad 18c.$$

In Equations 18a-b, $\xi_1$ and $\xi_2$ quantify the percentage of the input pixels that contain the spatial overlap with respect to $\theta_1(x,y)$ and $\theta_2(x,y)$, respectively, and the final $\xi$ is taken as the *max* of these two values.

For the digital implementation of the diffractive optical network training outlined above, we developed a custom-written code in Python (v3.6.5) and TensorFlow (v1.15.0, Google Inc.). The backpropagation updates were calculated using the Adam [27] optimizer with its parameters set to be the default values as defined by TensorFlow and kept identical in each model. The learning rate was set to be 0.001 for all the diffractive network models presented here. The training batch size was taken as 75 during the deep learning-based training of the presented diffractive networks. The training of a 5-layer diffractive optical network with 40K diffractive neurons per layer for 20,000 epochs takes approximately 24 days using a computer with a GeForce GTX 1080 Ti Graphical

Processing Unit (GPU, Nvidia Inc.) and Intel® Core ™ i7-8700 Central Processing Unit (CPU, Intel Inc.) with 64 GB of RAM, running Windows 10 operating system (Microsoft).

## Figures

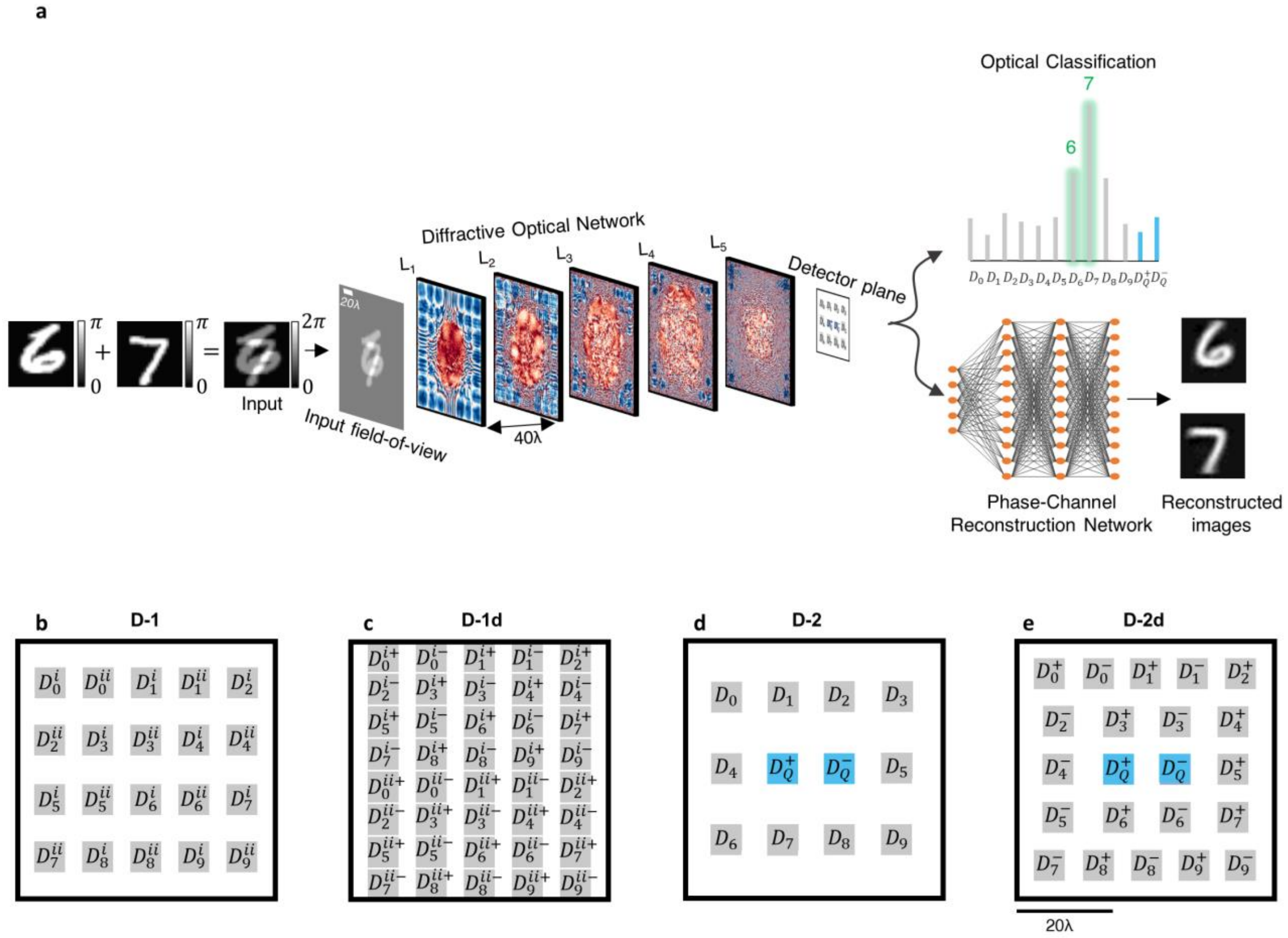

**Figure 1. Schematic of a diffractive optical network that can all-optically classify overlapping phase objects despite phase ambiguity at the input; this diffractive network also compresses the input spatial information at its output plane for simultaneous reconstruction of the individual phase images of the overlapping input objects using a back-end electronic neural network. a**, Optical layout of the presented 5-layer diffractive optical networks that can all-optically classify overlapping phase objects, e.g., phase-encoded handwritten digits, despite the phase ambiguity at the input plane due to spatial overlap. The diffractive optical network processes the incoming object waves created by the spatially overlapping, phase-encoded digits e.g., '6' and '7', to correctly reveal the classes of both input objects (green). A separately trained shallow electronic neural network (with 2 hidden layers) rapidly reconstructs the individual phase images of both input objects using the optical signals detected at the output plane of the diffractive network. **(b-d)**, Different detector configurations and class encoding schemes at the output plane of a diffractive network, devised to represent all the possible data class combinations at the input field-of-view created by overlapping phase objects.

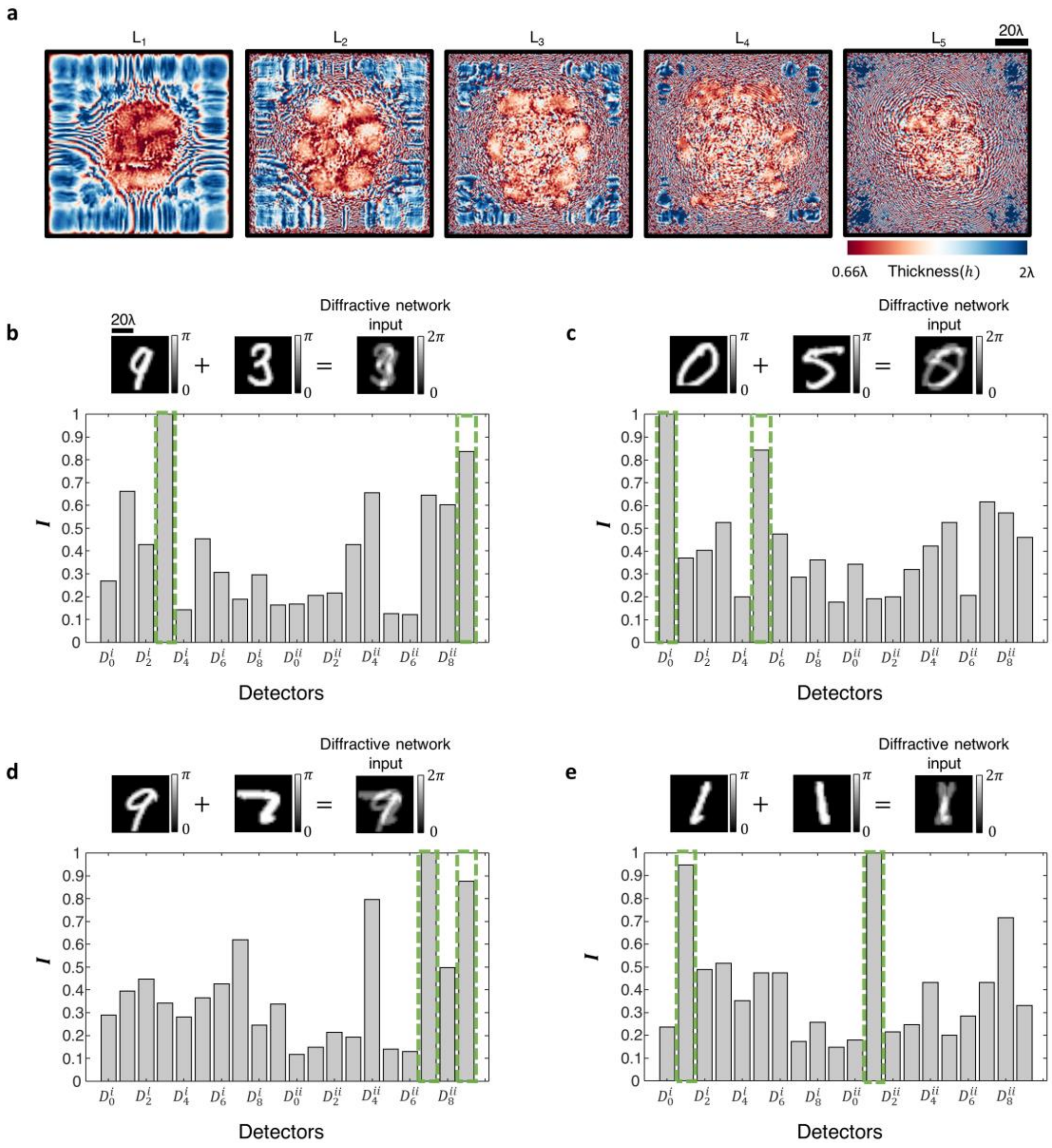

**Figure 2. All-optical classification of spatially-overlapping phase objects using the diffractive network D²NN-D1, based on the detector layout scheme (D-1) shown in Fig. 1b. a**, The thickness profiles of the diffractive layers constituting the diffractive network D²NN-D1 at the end of its training. This network achieves 82.70% blind inference accuracy on the test image set $T_2$. **b-e**, **Top:** Individual phase objects (examples) and the resulting input phase distribution created by their spatial overlap at the input field-of-view. **Bottom:** The normalized optical signals, $\boldsymbol{I}$, synthesized by D²NN-D1 at its output detectors. The output detectors with the largest 2 signals correctly reveal the classes of the overlapping input phase objects (indicated with the green rectangular frames).

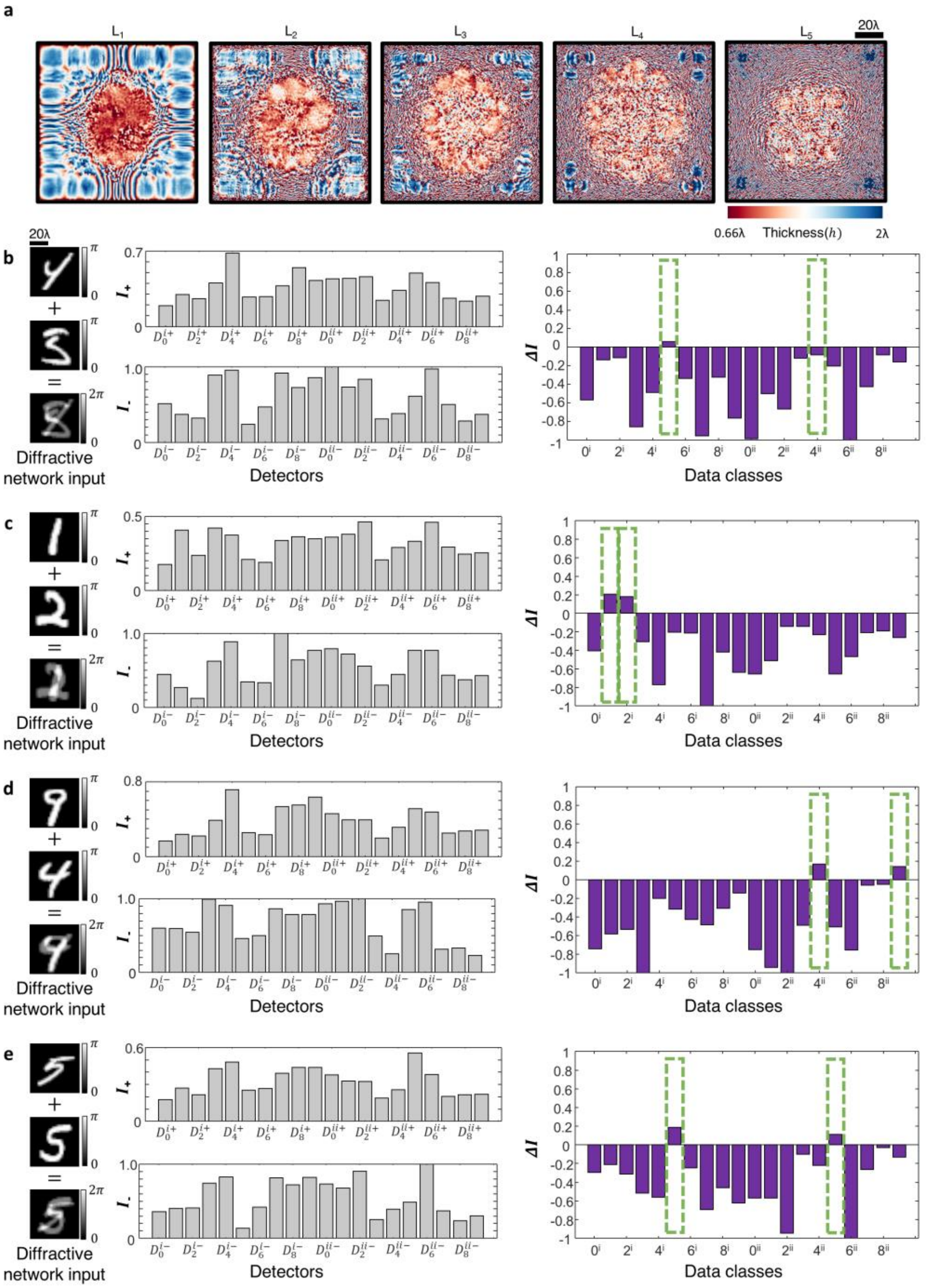

**Figure 3. All-optical classification of spatially-overlapping phase objects using the diffractive network D²NN-D1d, based on the detector layout scheme D-1d shown in Fig. 1c. a**, The thickness profiles of the diffractive layers constituting the diffractive network D²NN-D1d at the end of its training. This network achieves 85.82% blind inference accuracy on the test image set $T_2$. **b-e**, **Left:** Individual phase objects (examples) and the resulting input phase distribution created by their spatial overlap at the input field-of-view. **Middle:** The normalized optical signals, $\boldsymbol{I}_+$ and $\boldsymbol{I}_-$, synthesized by D²NN-D1d at its output detectors. **Right:** The resulting differential signal, $\Delta\boldsymbol{I} = \boldsymbol{I}_+ - \boldsymbol{I}_-$. The largest two differential optical signals correctly reveal the classes of the overlapping input phase objects (indicated with the green rectangular frames).

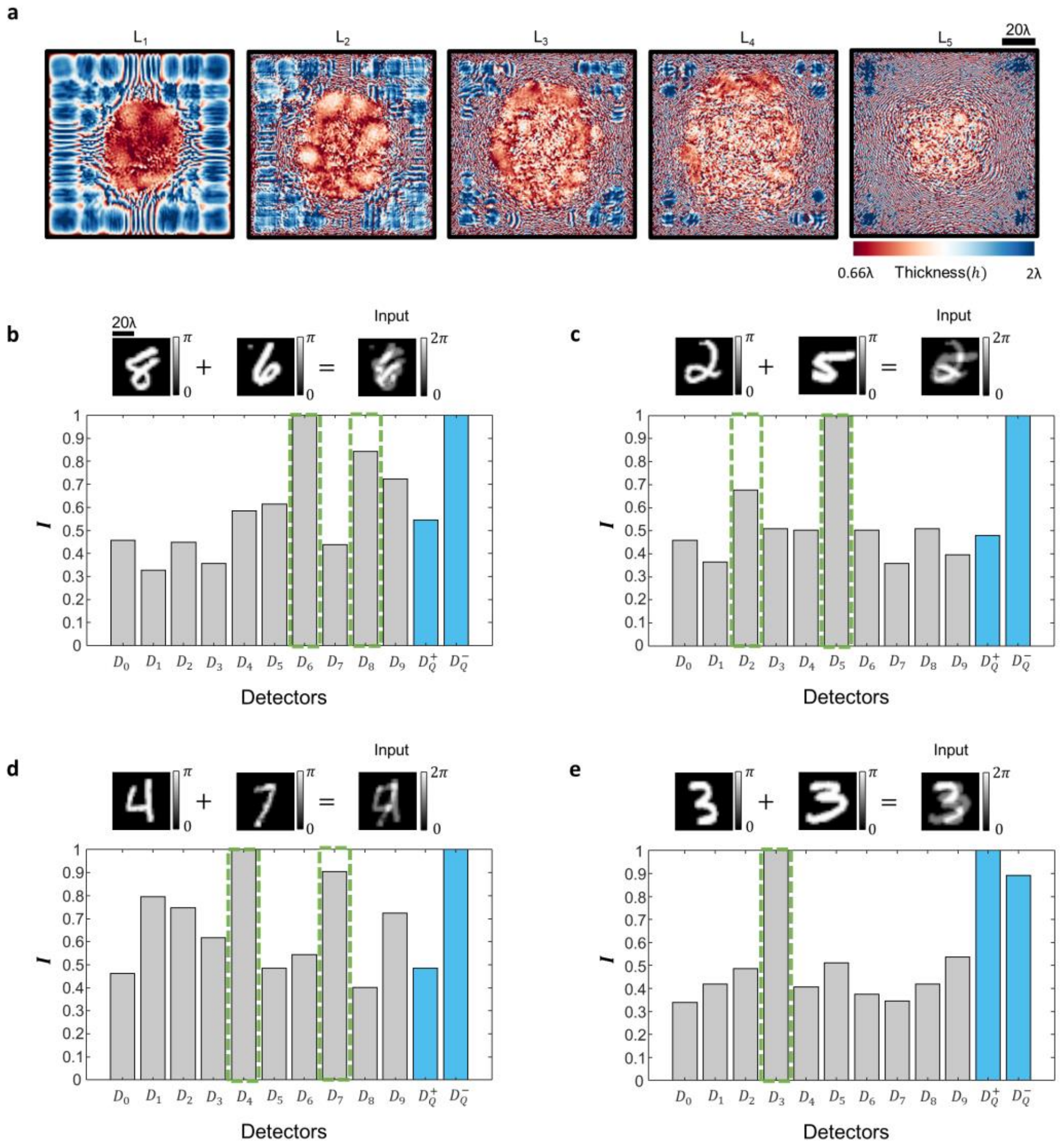

**Figure 4. All-optical classification of spatially-overlapping phase objects using the diffractive network D²NN-D2, based on the detector layout scheme D-2 shown in Fig. 1d. a**, The thickness profiles of the diffractive layers constituting the diffractive network D²NN-D2 at the end of its training. This network achieves 82.61% blind inference accuracy on the test image set $T_2$. **b-e**, **Top:** Individual phase objects (examples) and the resulting input phase distribution created by their spatial overlap at the input field-of-view. **Bottom:** The normalized optical signals, $\boldsymbol{I}$, synthesized by D²NN-D2 at its output detectors. For $I_{D_Q^+} < I_{D_Q^-}$, the largest two optical signals correctly reveal the classes of the overlapping input phase objects (indicated with the green rectangular frames in b,c,d). For $I_{D_Q^+} \geq I_{D_Q^-}$, the largest optical signal correctly reveals the class of the overlapping input phase objects (indicated with the green rectangular frame in e).

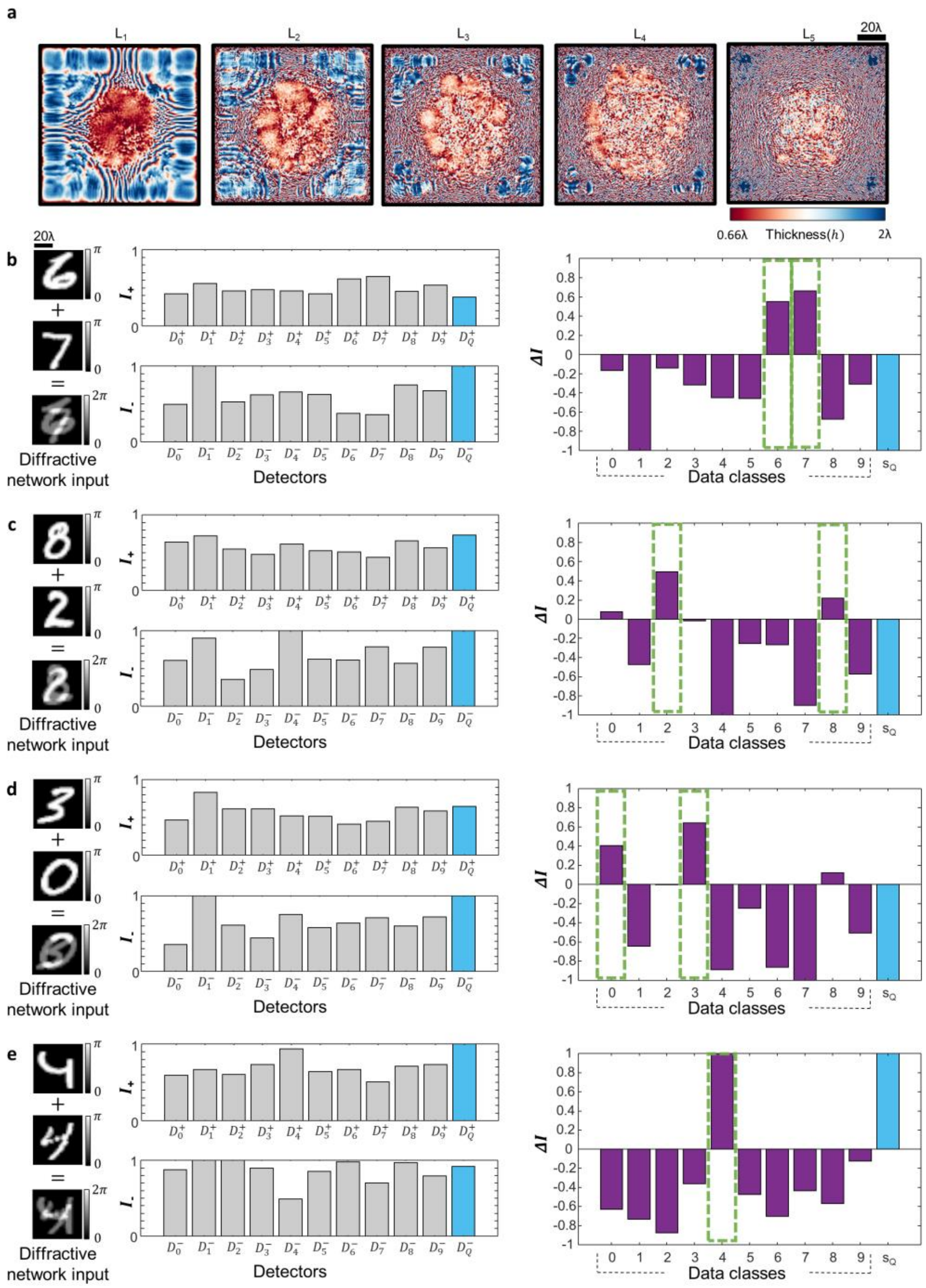

**Figure 5. All-optical classification of spatially-overlapping phase objects using the diffractive network D$^2$NN-D2d, based on the detector layout scheme D-2d shown in Fig. 1e. a**, The thickness profiles of the diffractive layers constituting the diffractive network D$^2$NN-D2d at the end of its training. This network achieves 85.22% blind inference accuracy on the test image set $T_2$. **b-e**, **Left:** Individual phase objects (examples) and the resulting input phase distribution created by their spatial overlap at the input field-of-view. **Middle:** The normalized optical signals, $\boldsymbol{I}_+$ and $\boldsymbol{I}_-$ , synthesized by D$^2$NN-D2d at its output detectors. **Right**: The differential optical signal, $\Delta\boldsymbol{I} = \boldsymbol{I}_+ - \boldsymbol{I}_-$ (purple). For $I_{D_Q^+} < I_{D_Q^-}$, the largest two differential optical signals correctly reveal the classes of the overlapping input phase objects (indicated with the green rectangular frames in b,c,d). For $I_{D_Q^+} \geq I_{D_Q^-}$, the largest differential optical signal correctly reveals the class of the overlapping input phase objects (indicated with the green rectangular frame in e).

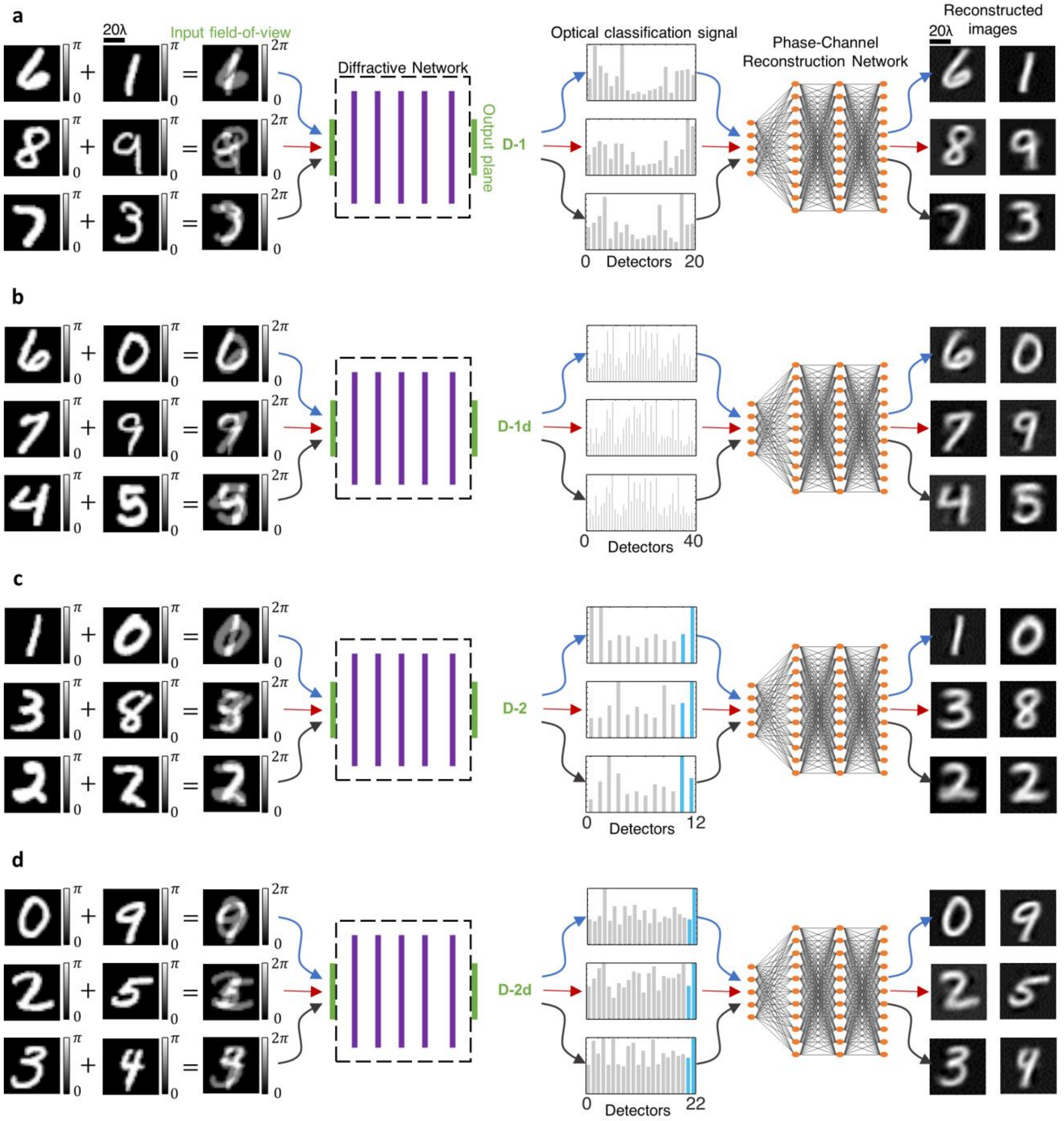

**Figure 6. Reconstruction of spatially overlapping phase images using a diffractive optical front-end (encoder) and a separately trained, shallow electronic neural network (decoder) with 2 hidden layers.** The front-end diffractive networks are (**a**) $D^2$NN-D1, (**b**) $D^2$NN-D1d, (**c**) $D^2$NN-D2, and (**d**) $D^2$NN-D2d, shown in Figs. 2a, 3a, 4a and 5a, respectively. The detector layouts at the output plane of these diffractive networks are (**a**) D-1, (**b**) D-1d, (**c**) D-2, and (**d**) D-2d with $2M$, $4M$, $M+2$ and $2M+2$ single pixel detectors as shown in Figs. 1b-d, respectively; for handwritten digits $M=10$. These four designs create a compression ratio of 39.2×, 19.6×, 65.33× and 35.63× between the input and output fields-of-view of the corresponding diffractive network, respectively. The mean SSIM and PSNR values achieved by these phase image reconstruction networks are depicted in Table 2 along with the corresponding standard deviation values computed over the 10K test input images ($T_2$).

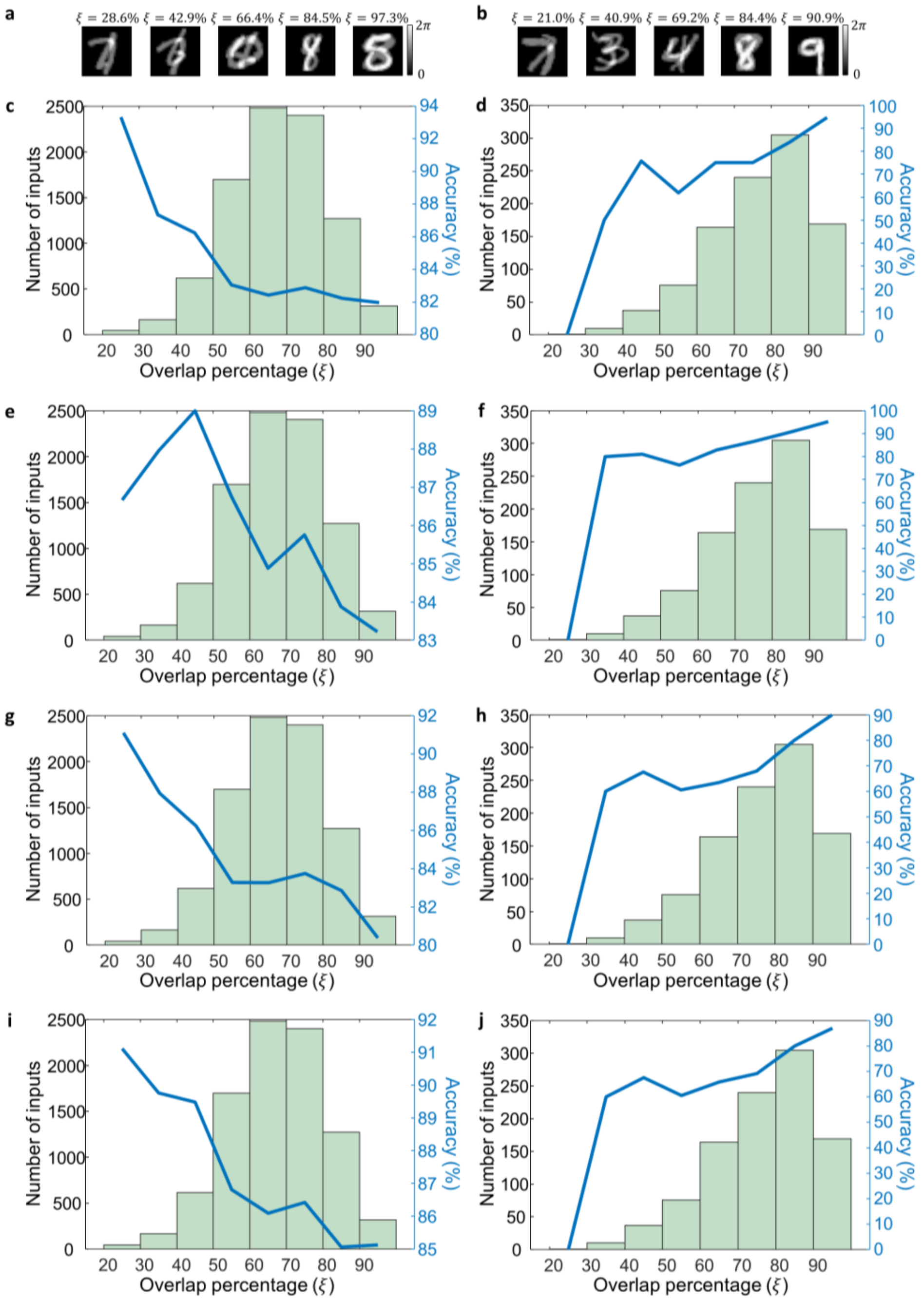

**Figure 7. The variation in the optical blind inference accuracies of the presented diffractive optical networks as a function of the spatial overlap percentage ($\xi$) between the two input phase objects. a,** Sample input images from the test set $T_2$ containing overlapping phase objects from different data classes along with the corresponding overlap percentages, $\xi$. **b,** Same as (**a**), except the overlapping objects are from the same data class. **c,** The blind inference accuracy of the diffractive network, $D^2$NN-D1, as a function of the overlap percentage, $\xi$, and the histogram of $\xi$, for test inputs in $T_2$ that contain phase objects from two different data classes. **d,** Same as (**c**), except that the test inputs contain phase objects from the same data class. **e** and **f,** Same as (**c** and **d**), except, the diffractive network design is $D^2$NN-D1d. **g** and **h,** Same as (**c** and **d**), except, the diffractive network design is $D^2$NN-D2. **i** and **j,** Same as (**c** and **d**), except, the diffractive network design is $D^2$NN-D2d.

# Tables

| | Test Set | |
|---|---|---|
| | $T_2$ 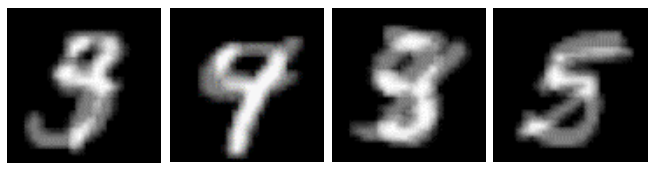 examples | $T_1$ 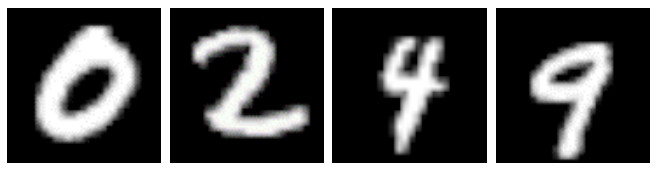 examples |
| Diffractive Network | Accuracy (%) | Accuracy (%) |
| $D^2$NN-D1 | 82.70 | 90.59 |
| $D^2$NN-D1d | 85.82 | 93.30 |
| $D^2$NN-D2 | 82.61 | 93.38 |
| $D^2$NN-D2d | 85.22 | 94.20 |

**Table 1**. The summary of the optical blind inference accuracies achieved by the presented diffractive optical networks on test sets $T_2$ and $T_1$ along with some input examples from these datasets.

| | | Optical Classification | Image reconstruction | |
|---|---|---|---|---|
| Diffractive Network | Number of detectors ($M$=10) | Accuracy on $T_2$ (%) | SSIM | PSNR (dB) |
| $D^2$NN-D1 | $2M$ | 82.70 | 0.52±0.12 | 15.09±2.32 |
| $D^2$NN-D1d | $4M$ | 85.82 | 0.57±0.10 | 16.02±2.21 |
| $D^2$NN-D2 | $M$+2 | 82.61 | 0.49±0.10 | 14.55±2.17 |
| $D^2$NN-D2d | $2M$+2 | 85.22 | 0.57±0.12 | 15.60±2.37 |

**Table 2.** The comparison of the presented diffractive optical networks, in terms of (1) all-optical overlapping object classification accuracies on $T_2$ and (2) the quality of the image reconstruction achieved through separately-trained, shallow, electronic networks (decoder). The mean SSIM and PSNR values and their standard deviations were computed over the entire 10K blind test inputs ($T_2$).